\documentclass[12pt,a4paper]{article}
\usepackage[utf8]{inputenc}
\usepackage[T1]{fontenc}
\usepackage{amsmath}
\usepackage{amsfonts}
\usepackage{authblk}
\usepackage{amssymb}
\usepackage{graphicx}
\newcommand{\be}{\begin{equation}}
\newcommand{\ee}{\end{equation}}
\newcommand{\bea}{\begin{eqnarray}}
\newcommand{\eea}{\end{eqnarray}}

\title{\vspace{-1.8in} Thermal Equilibrium in String Theory \\ in the Hagedorn Phase}
\author{Ram Brustein, Yoav Zigdon}
\affil{{\normalsize Department of Physics, Ben-Gurion University, Beer-Sheva 84105, Israel} \\
	{\small\tt ramyb@bgu.ac.il\ \ \ \ yoavzig@post.bgu.ac.il}}
\date{}
\begin{document}
	\maketitle
	\begin{abstract}
In string theory, a thermal state is described by compactifying Euclidean time on a thermal circle $S^{1}_{\beta}$, of fixed circumference. However, this circumference is a dynamical field which could vary in space, therefore thermal equilibrium is not guaranteed. We discuss a thermal state of type II string theory near and above the Hagedorn temperature and show that the circumference of the thermal circle can indeed be fixed and stabilized in the presence of a uniform isotropic flux.
 We solve the equations of motion derived from an action that reproduces the tree-level string S-matrix. We find solutions with the topologies of $S^{1}_{\beta}\times S^2 \times {\cal M}^{d-2}$ at a fixed temperature, which include a space-filling winding-mode condensate and a uniform Neveu-Schwarz Neveu-Schwarz flux supported on $S^1 _{\beta}\times S^2$. The solutions that we find have either a linear dilaton or a constant dilaton, in which case, we find solutions with either a cosmological constant or a Ramond-Ramond flux.  We then compare our solutions to the cigar and cylinder backgrounds associated with the $SL(2,R)/U(1)$ coset theory, which include a winding-mode condensate but without flux. We also compare and contrast our solutions with the non-uniform Horowitz-Polchinski solution, which also possesses a winding-mode condensate and is characterized by an approximate thermal equilibrium near the Hagedorn temperature.
	\end{abstract}
\newpage
	\tableofcontents
	
	\section{Introduction}

We discuss thermal equilibrium of (super)strings near and above the Hagedorn temperature -- the Hagedorn phase of string theory.

In quantum field theories without gravity, one describes a thermal state by compactifying Euclidean time on a circle, the thermal circle $S^{1}_{\beta}$, and imposing appropriate boundary conditions for bosons and fermions. The circumference of the thermal circle is equal to the inverse temperature in units in which the Boltzmann constant and $\hbar$ are set to one.  However, in theories containing gravity, such as string theory, the size of the thermal circle becomes a dynamical field, the ``radion.'' In this case,  thermal equilibrium, corresponding to a thermal circle of constant circumference is not guaranteed to exist. Furthermore, the Jeans instability plagues asymptotically flat thermal backgrounds \cite{Jeans}.

All of the string theories feature a special temperature, the ``Hagedorn temperature'', above which the one-loop free energy of a free string diverges \cite{Sundborg},\cite{Alvarez},\cite{Tye},\cite{Sathipalan},\cite{Obrien},\cite{AW}. Near and below this temperature, closed string modes that wind once around the thermal circle become light and can be described in terms of two conjugate complex scalar fields $\chi$ and $\chi^*$. The field $\chi$ is known as the thermal scalar.  Above the Hagedorn temperature, $T_H$, these fields become tachyonic (for constant dilaton), thus the existence and stability of a high-temperature phase may be hindered.

A system of free closed strings slightly below $T_H$ can be described by highly-excited, long strings which can be viewed as performing random walks in target space, with a step size of the order of the string length \cite{SS}. The energy, entropy and length of these strings all scale linearly with the number of random-walk steps. Near $T_H$, one can expect a transition between a phase of long strings and that of short strings. A connection between the effective description in terms of the thermal scalar and the excited string states was established by evaluating the energy and density of states \cite{BV}, \cite{HP}.

For interacting closed strings near $T_H$, it was demonstrated in \cite{AW} that the effective potential of the winding modes includes an important non-local quartic term arising from the interaction of $\chi$ and $\chi^*$ with the radion. By including this interaction, Horowitz and Polchinski (HP) \cite{HP} were able to show the existence of an approximate thermal equilibrium slightly below the Hagedorn temperature. They considered a winding-mode condensate backreacting on an otherwise constant dilaton and $S^{1}_{\beta}\times \mathbb{R}^d$ geometry. The geometry of this backreacted solution in the compact and radial directions can be visualized as a cylinder whose circumference shrinks by a small amount in the region of space where the winding-mode condensate is localized, as depicted in the middle of Fig. 1. Based on the scaling properties of the entropy and temperature of strings and black holes (BHs), it was speculated that as the string coupling varies, the two objects can transform  into each other \cite{Bowick:1985af},\cite{Susskind:1993ws},\cite{HP0},\cite{DV}. Recently, the HP solution and the possibility that it continuously transits into a BH were discussed in \cite{MaldacenaLargeD},\cite{CMW}.

In a previous article \cite{HagedornEFT}, we derived a low-energy effective field theory (EFT) for bosonic and type II closed strings in the Hagedorn phase. The action of the EFT was calculated from the requirement that its amplitudes are identical to the string S-matrix elements. One of the main differences between our EFT action and the HP action is the additional local quartic interaction term of the winding modes.

This quartic term can be interpreted as coming from strings that interact at their intersections.  The strength of the interaction is proportional to the number density of the strings squared.  Since the entropy density of winding strings is proportional to the winding-mode condensate squared \cite{CFT4dS},\cite{MaldacenaLargeD}, the local quartic term is proportional to the entropy density squared \cite{CFT4dS},\cite{dSHigh},\cite{emerge},\cite{strungout}. Provided that the entropy density scales with the number density of the strings, a scaling with the number density squared is obtained.  Not coincidentally, the strings free energy is formally similar to those appearing in the literature on interacting polymers ({\em e.g.}, \cite{polytext}).

In this paper we establish the existence of a state of thermal equilibrium of strings in the Hagedorn phase by showing that the radion can be stabilized in a space with the geometry of $S^{1}_{\beta}\times S^2 \times {\cal M}^{d-2}$, and in the presence of a specific three-form flux.
The flux produces an outward force on the thermal circle which exactly counterbalances the inward gravitational force of the winding modes. The dilaton in our constructions is either linear in one of the coordinates, or constant. In the latter case, either a Ramond-Ramond (RR) flux or a cosmological constant (CC) are added to the effective action. The backreaction of the RR flux produces an Anti-de-Sitter (AdS) space component in the resulting product geometry, while the CC leads to a flat-space component.

	\begin{figure}[t]
\vspace{-1.5in}
		\begin{center}
		\includegraphics[scale=0.5]{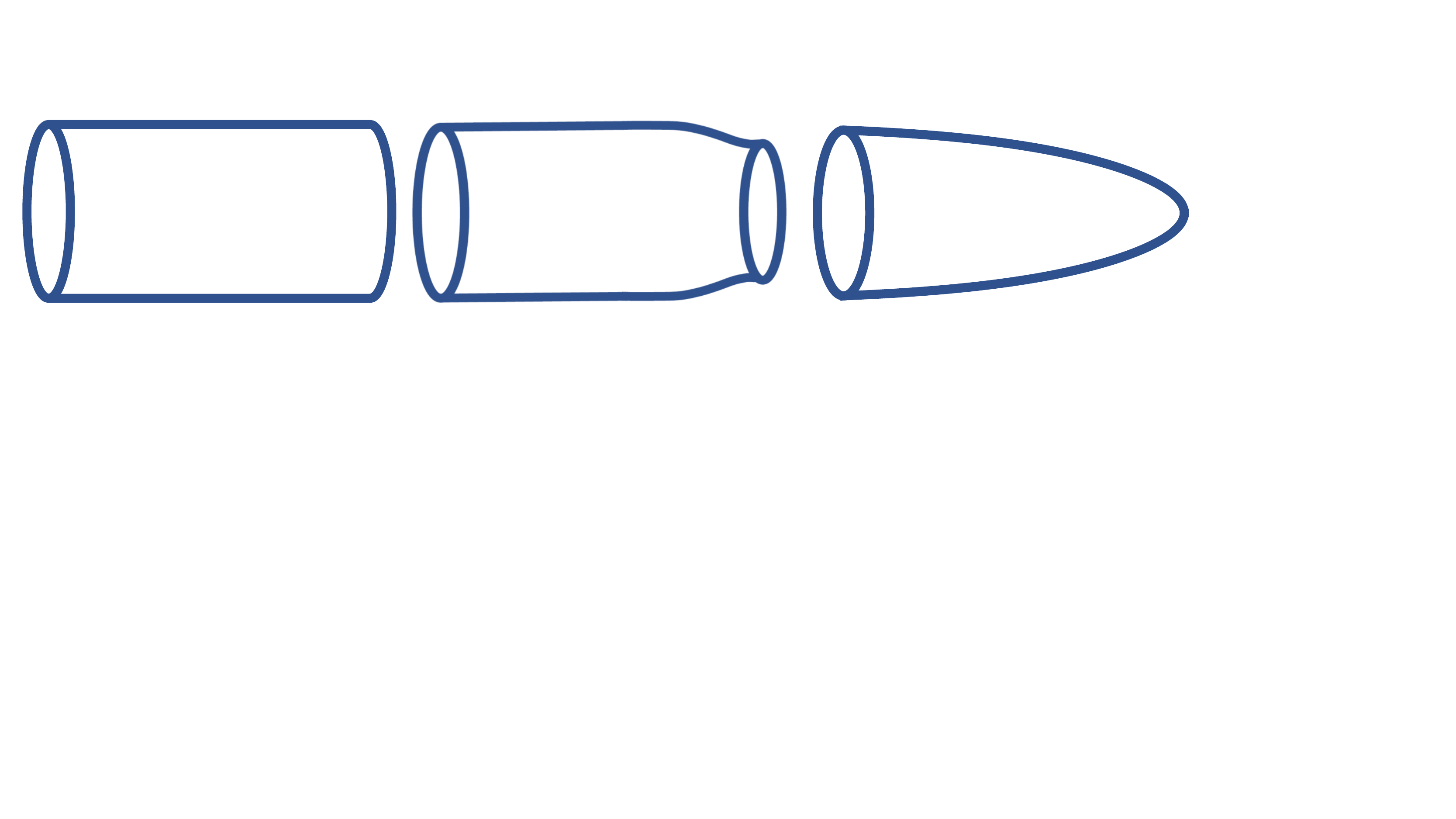}
		\end{center}
\vspace{-2.5in}
		\caption{ Shown from left to right are the compact time-radial parts of the geometries of our thermal equilibrium solutions, of the HP solution and of the cigar background. The three-form flux stabilizes the radius of the thermal circle, counteracting gravity. Without flux, an approximate thermal equilibrium is described by the HP solution. The cigar background deviates significantly from  thermal equilibrium: The thermal circle is contractible and vanishes at the horizon. }
	\end{figure}

The terminology ``thermal equilibrium'' admits at least two interpretations. One is that the local temperature is constant in space, as for the solutions written in the paper. The second is that there is no net energy flow between components of the system, for example, a black hole and its emitted Hawking radiation. When the gravitational potential has a non-trivial profile, the former interpretation fails; the local temperature depends on the position. We view the solutions as interesting examples that are similar to quantum mechanical and quantum field theory systems without gravity that exhibit thermal equilibrium.

Euclidean BH string backgrounds exhibit significant deviations from thermal equilibrium. The thermal circle at infinity shrinks towards a ``tip'' corresponding to the BH horizon. Thus, the local temperature varies in Euclidean space, increasing towards the tip. 
  As in the HP solution, the BH solutions include a winding-mode condensate \cite{Dabholkar},\cite{Kutasov1},\cite{Sunny1},\cite{Sunny2}. A specific example of such a background is a 2D Euclidean solution described by the $SL(2,R)/U(1)$ coset theory  \cite{cigarRef1},\cite{EdCigar},\cite{DVV}. This is one of a few backgrounds whose worldsheet conformal field theory (CFT) is known to be perturbatively exact in $\alpha'$. A 10D solution containing the cigar is obtained by considering the near-horizon region of $k$ near-extremal NS5 branes \cite{MaldacenaNS5}. Recently, a solution of this cigar taking into account the backreaction of the winding-mode condensate was found \cite{WindingBackreaction}. See also \cite{Mertens1}, \cite{SunnyNew},\cite{MaldacenaLargeD},\cite{EntropyCondensate},\cite{Amit},\cite{CMW}.

All the backgrounds discussed above feature a winding-mode condensate, however, they differ in some important aspects. For instance, Euclidean BHs have a contractible $S^{1} _{\beta}$, while the HP solution and our solutions do not. The HP solution has asymptotic temperature which is near and below the Hagedorn temperature, whereas our solutions have flux and a fixed $S^{1} _{\beta}$ with temperature near but above the Hagedorn temperature. Another important difference is that neither the HP solution nor the $SL(2,R)/U(1)$ cigar and cylinder backgrounds are uniform in space, while our solutions (with a constant dilaton) are uniform. Additional similarities and differences are discussed later in the paper. The geometries of the different solutions are depicted in Fig. 1.

The paper is organized as follows. In the next section we briefly review the EFT for the light winding modes near the Hagedorn temperature \cite{HagedornEFT}. In Section 3 we write the action for type II light closed string modes with fluxes. We then derive equations of motion (EOM) and solve them explicitly. Section 4 is devoted to a proposal that some worldsheet field theories flow to a fixed point, which correspond to the target-space solutions (without RR flux) of the EFT. In Sections 5 and 6 we compare and contrast the $SL(2,R)/U(1)$ cigar and cylinder backgrounds, the HP background and our thermal equilibrium backgrounds. Conclusions appear in Section 7.

    \section{Effective Action for Strings near the Hagedorn Temperature}

In this section we review the effective action of type II closed strings in thermal equilibrium in the Hagedorn phase which was derived in \cite{HagedornEFT}.  There, the compact-compact component of the metric was fixed to unity by enforcing thermal equilibrium. This constraint will be relaxed in the next section. A similar EFT for the heterotic string was derived in \cite{Troost}.

    We adopt the following notations\footnote{In \cite{HagedornEFT}, the winding-mode fields were denoted by $\phi,\phi^*$ and the d-dimensional dilaton by $\psi$.}: $\chi$ and $\chi^*$ denote fields representing the winding modes that wind once around the thermal circle. The metric in the $d$ spatial dimensions is denoted by $G_{\mu \nu}$ and the $d$-dimensional dilaton $\Phi_d$. Other modes such as the Kalb-Ramond two-form and the graviphoton are set to zero (up to  a gauge) in this section.

    The Hagedorn temperature of type II superstring theory is given by \cite{Sundborg}:
    \begin{equation}
    T_H = \frac{1}{2\pi \sqrt{2\alpha'}}.
    \end{equation}
    The effective action of the winding modes in type II string theory near the Hagedorn temperature, following from S-matrix calculations, is given by
    \begin{equation}\label{EFT1}
    S_1 = \beta\int d^{d} x \sqrt{G}e^{-2\Phi_d}\left(G ^{\mu \nu}\partial_{\mu}\chi \partial_{\nu}\chi^* + m^2 _{\chi}\chi \chi^*+2\frac{ \kappa^2}{\alpha'}(\chi \chi^*)^2\right),
    \end{equation}
    where $\kappa^2$ is related to the Newton constant, $\kappa^2 = 8\pi G_N$. The quartic interaction term is one of the main differences between our effective action and the HP effective action. To compute it, we subtracted exchange amplitudes\footnote{These include four ingoing and outgoing string states, with an on-shell string mode, which can either be a dilaton, a graviton, a radion or a gauge boson.} in the EFT from a four-point worldsheet correlation function of on-shell winding vertex operators \cite{HagedornEFT}. In \cite{Klebanov}, the quartic coupling was computed up to a multiplicative constant, and the authors argued that the  quartic coupling is positive after performing a field redefinition such that the three-point coupling of the radion with $\chi,\chi^*$ has no derivatives. In \cite{Troost}, the quartic coupling for the heterotic string was calculated and found to be $+3\; \frac{\kappa^2}{\alpha'}$.

    The mass squared of the winding modes is given by
    \begin{equation}\label{mphi2}
    m^2 _{\chi} = -\frac{2}{\alpha'} + \frac{1}{4\pi^2 T^2 (\alpha')^2}.
    \end{equation}

    The mass-squared in Eq.~(\ref{mphi2}) and the quartic-self interaction can be expressed as follows:
    \begin{eqnarray}
    &m_{\chi}^2 = -2c_2\epsilon T^2+O(\epsilon^2),\\
    &\frac{2\kappa^2}{\alpha'} = c_2 \kappa^2 T^2+O(\epsilon),\\
    & c_2=16\pi^2,
    \end{eqnarray}
    with
    \begin{equation}\label{epsilon}
    \epsilon\equiv  \frac{T-T_H}{T}.
    \end{equation}
    Substituting these expressions into $S_1$, one finds, to leading order in $\epsilon$,
    \begin{eqnarray}
    &S_1 = \beta\int d^{d} x \sqrt{G}e^{-2\Phi_d}\left\{G ^{\mu \nu}\partial_{\mu}\chi \partial_{\nu}\chi^* - 2c_2 \epsilon T^2\chi \chi^*+c_2 \kappa^2 T^2(\chi \chi^*)^2\right\}.
    \end{eqnarray}

    We can add the standard action for the graviton and dilaton to the action for the winding modes,
    \begin{eqnarray}
    &S_2 = -\frac{\beta}{2\kappa^2} \int d^{d} x \sqrt{G} e^{-2\Phi_d} \biggl\{R-2\Lambda+4G^{\mu \nu}\partial_{\mu} \Phi_d \partial_{\nu} \Phi_d\biggr\},
    \end{eqnarray}
    where $\Lambda$ is a cosmological constant (CC), which is not a free parameter in the theory. However, in the next section we will add a fine-tuned CC in order to find a simple flat-space,  constant-dilaton solution of the EFT. For other solutions we will set $\Lambda=0$. Note also that we have not added a kinetic term for $\sigma$ because we assume here that this field is constant. We will add this term in the next section.

    The total EFT action is therefore given by the sum of $S_1$ and $S_2$,
    \begin{eqnarray}
    S &=& \beta\int d^{d} x \sqrt{G}e^{-2\Phi_d}\Biggl\{-\frac{1}{2 \kappa^2}R+\frac{1}{\kappa^2}\Lambda- \frac{2}{\kappa^2}G^{\mu \nu}\partial_{\mu}\Phi_d\partial_{\nu} \Phi_d  \cr &+&  G ^{\mu \nu}\partial_{\mu} \chi \partial_{\nu}\chi^* - 2c_2 \epsilon T^2\chi \chi^*+c_2 \kappa^2 T^2(\chi \chi^*)^2\Biggr\}.
    \label{totalEFT}
    \end{eqnarray}
    In principle, more terms involving products of $\chi,\chi^*$ and their derivatives should be added; however, for solutions in which $\chi,\chi^*$ are constant and scale with a positive power of $\epsilon$, which are the class of solutions that we will consider, it is justified to ignore such terms \cite{CFT4dS}. 	

\section{Flux Compactification Backgrounds}

    We wish to demonstrate the existence of solutions of the EOM of the light fields, in which the size of the thermal circle is fixed. The winding strings attract each other due to gravity, but also produce flux that keeps the thermal circle from collapsing. This can be viewed as an effective model for a thermal bath of strings at a temperature slightly above the Hagedorn temperature.

    For the string frame $\tau-\tau$ metric $G_{\tau \tau}$, we use the notation from \cite{PolchinskiI},\cite{PolchinskiII},
    \begin{equation}\label{sigma}
    G_{\tau \tau}=e^{2\sigma}.
    \end{equation}

    The effective action for the winding modes is slightly changed relative to Eq.~(\ref{EFT1}):
    \begin{eqnarray}\label{StartingPoint}
    &S_{\chi,\chi^*}= \beta  \int d^{d} x \sqrt{G}e^{-2\Phi_d}\left(G ^{\mu \nu}\partial_{\mu}\chi \partial_{\nu}\chi^* + \frac{\beta^2 e^{2\sigma}-\beta_H ^2}{4\pi^2 (\alpha')^2}\chi \chi^*+\frac{2\kappa^2}{\alpha'} (\chi \chi^*)^2\right).
    \end{eqnarray}
     The new factor $e^{2\sigma}$ originates from the interaction of the winding modes with the radion $\sigma$ \cite{HP},\cite{HagedornEFT}.

Next, we add the Neveu-Schwarz Neveu-Schwarz (NS-NS) sector action for the light fields,
\begin{eqnarray}\label{NSNS}
&S_{NS-NS} = -\frac{\beta}{2\kappa^2} \int d^{d} x \sqrt{G} e^{-2\Phi_d} \left(R-2\Lambda+4G^{\mu \nu}\partial_{\mu} \Phi_d \partial_{\nu} \Phi_d+\right. \nonumber \\
&\left.- G^{\mu \nu}\partial_{\mu}\sigma\partial_{\nu}\sigma-\frac{1}{4}e^{-2\sigma}H_{\tau \mu \nu}H^{~\mu \nu} _{\tau}\right). \ \ \
\end{eqnarray}
Here,
\begin{eqnarray}
H_{\tau \mu \nu}=\partial_{\nu} B_{\tau \mu}- \partial_{\mu} B_{\tau \nu}.
\end{eqnarray}
We set the graviphoton to zero (up to a gauge). We also set to zero the spatial elements of the Kalb-Ramond field strength $H_{\mu \nu \lambda}=0$.

The two-form $B$ with one $\tau$-leg $B_{\tau \mu}$,  couples to the $\chi$ current,
\begin{equation}\label{Bchichi}
	i\frac{\beta}{2\pi \alpha'}\beta \int d^d x~ e^{-2\Phi} \sqrt{G} B_{\tau} ^{~\mu}\left(\chi \partial_{\mu} \chi ^* - \chi^* \partial_{\mu} \chi\right) . 
\end{equation}
This is reproduced from an amplitude associated with two winding modes of winding numbers $\pm 1$ and the B-field \cite{HagedornEFT}, as explained in the appendix. 
This additional term renders the kinetic term of $\chi$ having covariant derivatives, but for simplicity of notation we keep the $\partial$ notation.

Similarly, we consider a RR potential $(p-1)$-form $C_{p-1}$, with one leg on $\tau$, which is coupled to a  Euclidean $(p-2)$-D-brane. We will solve the EOM outside any sources.
The RR action for the field strength $F_p = dC_{p-1}$, with one leg in $\tau$, is given by
\begin{equation}\label{SRR}
S_{RR} = \frac{\beta}{4\kappa^2 (p-1)!} \int d^d x \sqrt{G} e^{-\sigma} F_{\tau \mu_2...\mu_p} F_{\tau} ^{~\mu_2...\mu_p}.
\end{equation}
Below, we list the EOM derived from the action\footnote{In type IIB, Eq. (\ref{SRR}) has an additional factor of a half. Also, note that the Chern-Simons term involving the product of the potentials and field strengths, vanishes in our case because they all have a leg in $\tau$.}
\begin{equation}
S=S_{\chi,\chi^*}+S_{NS-NS} +S_{RR}.
\end{equation}

\subsection{Equations of motion}

\begin{itemize}

	\item
 {\em The $\chi$ and $\chi^*$ equations}
\begin{eqnarray}
\label{phiEOM}
&\frac{e^{2\Phi_d}}{\sqrt{G}}\partial_{\mu}\left(e^{-2\Phi_d}\sqrt{G}G^{\mu \nu} \partial_{\nu}\chi \right) =
&\frac{\beta^2 e^{2\sigma}-\beta_H ^2}{4\pi^2 (\alpha')^2} \chi +\frac{4\kappa^2}{\alpha'}(\chi)^2 \chi^*.
\end{eqnarray}
\begin{eqnarray}
&\frac{e^{2\Phi_d}}{\sqrt{G}}\partial_{\mu}\left(e^{-2\Phi_d}\sqrt{G}G^{\mu \nu} \partial_{\nu}\chi^* \right) =
&\frac{\beta^2 e^{2\sigma}-\beta_H ^2}{4\pi^2 (\alpha')^2} \chi^* +\frac{4\kappa^2}{\alpha'} (\chi^*)^2 \chi.
\end{eqnarray}

\item
{\em The Kalb-Ramond equation}
\begin{equation}
\label{TwoForm}
\frac{1}{\kappa^2}\partial_{\alpha} \left(\sqrt{G} e^{-2\Phi_d-2\sigma} H_{\tau} ^{~\alpha \mu} \right)=0.
\end{equation}

\item
{\em Ramond-Ramond equation}
\begin{equation}
\label{TwoFormR}
\frac{1}{\kappa^2}\partial_{\alpha} \left(\sqrt{G} e^{-\sigma} F_{\tau} ^{~\alpha \mu_3...\mu_p} \right)=0.
\end{equation}

\item
{\em The $\sigma$ equation}
\begin{eqnarray}
\label{SigmaR}
\frac{e^{2\Phi_d}}{\sqrt{G}}\partial_{\mu} \left( e^{-2\Phi_d}\sqrt{G}\partial^{\mu}\sigma\right) &= & -\frac{1}{4}e^{-2\sigma}H_{\tau \mu \nu}H^{~\mu \nu} _{\tau}  \cr -\frac{e^{2\Phi_d-\sigma}}{4(p-1)!}F_{\tau \mu_2 ...\mu_p}F_{\tau} ^{~\mu_2...\mu_p} &+& \frac{\beta^2\kappa^2}{2\pi^2 (\alpha')^2}\chi \chi^*e^{2\sigma}.
\end{eqnarray}

\item
{\em A linear combination of the spatial metric and dilaton equations}
\begin{eqnarray}
\label{EinsteinR}
&&R_{\mu \nu}-\partial_{\mu} \sigma \partial_{\nu} \sigma +2\nabla_{\mu}  \nabla_{\nu} \Phi_d -\frac{1}{2}H_{ \mu \lambda\tau}H_{\nu}^{~~ \lambda\tau}  -\frac{e^{2\Phi_d}}{2(p-2)!}F_{  \mu \tau \mu_3...\mu_p}F_{\nu}^{~~ \tau \mu_3 ...\mu_p} \cr && +\frac{e^{2\Phi_d}}{4(p-1)!} G_{\mu \nu} F_{\tau \mu_2...\mu_p} F^{\tau \mu_2 ... \mu_p } =2\kappa^2\partial_{\mu} \chi \partial_{\nu} \chi^*.
\end{eqnarray}

\item
{\em The dilaton equation}
\begin{eqnarray}
\label{dilatonR}
&&R -2 \Lambda + 4 \nabla^2\Phi_d -4\partial^{\mu} \Phi_d \partial_{\mu} \Phi_d-\partial_{\mu}\sigma \partial^{\mu}\sigma  -\frac{1}{4}H_{\tau \mu \nu}H^{\tau \mu \nu}\cr &&  = 2\kappa^2 \left[ \partial ^{\mu} \chi \partial_{\mu} \chi^* +\frac{\beta^2 e^{2\sigma}-\beta_H ^2}{4\pi^2 (\alpha')^2}\chi \chi^*+\frac{2\kappa^2}{\alpha'}(\chi \chi^*)^2\right].
\end{eqnarray}

\end{itemize}

\subsection{Solutions with a vanishing Ramond-Ramond flux}

\begin{itemize}
	\item
	{\em The $\chi$ solution}
	
Consider $\beta = \beta_H = 2\sqrt{2}\pi \sqrt{\alpha'}$ and a small negative constant $\sigma=-\epsilon$. Since $\beta e^{\sigma}$ is the inverse temperature $T^{-1}$, one has
\begin{equation}
\epsilon = \frac{T-T_H}{T_H}.
\end{equation}
Setting these $\beta,\sigma$ values in Eq.~(\ref{phiEOM}), we find a constant solution for $\chi$, $\chi^*$,
\begin{equation}
\label{condensate2}
\chi \chi^* =\frac{\epsilon}{\kappa^2}.
\end{equation}
We are interested in the case $\epsilon\ll 1$ when $\chi,\chi^*$ are light and higher-order corrections to the effective action in Eq. (\ref{StartingPoint}) are suppressed.
Eq. (\ref{Bchichi}) implies a phase for $\chi$,
\begin{equation}
	\chi = \frac{\sqrt{\epsilon}}{\kappa} e^{-i\frac{\beta}{2\pi \alpha'}\int d x^{\mu}  ~B_{\tau \mu}},
\end{equation}
because we require that the covariant derivative of $\chi$ vanishes. 
\item
{\em The $\sigma$ solution}

Equation~(\ref{SigmaR}) has a constant solution, $|\sigma| =\epsilon\ll 1 $,
\begin{equation}
\label{SigmaEOM}
-\frac{1}{4}e^{-2\sigma}H_{\tau \mu \nu}H^{~\mu \nu} _{\tau}+\frac{4e^{2\sigma}\epsilon}{\alpha'}=0, 
\end{equation}
provided that
\begin{equation}
\label{Hsquared}
H_{\tau \mu \nu}H^{~\mu\nu} _{\tau}=\frac{16\epsilon}{\alpha'} +O(\epsilon^2) = 8c_2 \epsilon T^2+O(\epsilon^2).
\end{equation}

\item
{\em The Kalb-Ramond solution}

Here we consider flux supported on $S^1 _{\beta} \times S^2$, away from any possible stringy source,
\begin{equation}
H_{\tau i j} = h \epsilon_{\tau i j} ^{S^2},
\end{equation}
The indices $i$, $j=1,2$ correspond to directions in $S^2$.
Recall that
$
\epsilon_{\tau i j} \epsilon ^{\tau i j} =2.
$
From Eq. (\ref{Hsquared})  it follows that
\begin{equation}
\label{h2}
h^2 = \frac{8\epsilon}{\alpha'}.
\end{equation}
Below we will quantize this flux.

\item
{\em The spatial metric solution}

The Ricci tensor of a two-sphere of radius $r_0$ is given by
\begin{equation}
R_{i j}^{S^2} = \frac{1}{r_0^2}G_{i j} ^{S^2},
\end{equation}
where the metric $G_{ij}$ is the metric of a unit two-sphere,
while the Ricci scalar is given by
\begin{equation}
\label{Ricci}
R=\frac{2}{r_0 ^2}.
\end{equation}

Since
$
\epsilon_{\tau i j } \epsilon^{\tau i} _{~~k} = G_{j k},
$
Eq. (\ref{EinsteinR}) implies that
\begin{eqnarray}
\label{Sphere2a}
\frac{1}{r_0 ^2}G_{i j}&=& \frac{h^2}{2} G_{i j},\ i,~j=1,2 \\
G_{\mu\nu}&=&\delta_{\mu\nu},~\ \ \ \ \mu,~\nu\ne 1,2.
\end{eqnarray}
Therefore
\begin{equation}
\label{Sphere2b}
r_0 = \frac{1}{2}\frac{\sqrt{\alpha'}}{\sqrt{\epsilon}}.
\end{equation}
For a small $\epsilon$, this radius is parametrically larger than the string length.
The quantization of the flux dictates that $ e^{\frac{i}{2\pi \alpha'}\int_{S^1 \times S^2} H} =1$. It follows that
\begin{equation}
\epsilon = \frac{4}{n^2}~,~ n\in \mathbb{Z}.
\end{equation}
For $\epsilon\ll 1$, $n\gg 1$ and we can treat $\epsilon$ as a continuous parameter.

\end{itemize}

\subsubsection{A solution with the geometry of $S_{\beta} ^1 \times S^2\times \mathbb{R}^{d-2}$, a constant dilaton and a cosmological constant}

Here we consider adding a fine-tuned CC in order to find a flat-space solution with a constant dilaton,
\begin{equation}
\Phi_d (x) = \Phi_0,
\end{equation}
with $\Phi_0$ corresponding to a small string coupling.
This fixes the CC, as Eq.~(\ref{dilatonR}) implies that
\begin{equation}
\label{Dilaton2}
\frac{2}{r_0 ^2}-2 \Lambda- \frac{1}{2}h^2 = -\frac{4}{\alpha'}\epsilon^2.
\end{equation}
From Eqs.~(\ref{Sphere2b}), (\ref{h2}), it follows that
\begin{equation}\label{CC}
\Lambda= \frac{2\epsilon}{\alpha'} +O(\epsilon^2).
\end{equation}
Strictly speaking, in superstring theory, it is not allowed to add an arbitrary CC, yet it may be the case that the CC in Eq. (\ref{CC}) corresponds to the value of the minimum of the potential of some other target-space field. We have not identified which field could give rise to this positive CC. Note that in spite of the fact that the CC is positive, the $(d-2)$-dimensional space is flat. In what follows, we will set $\Lambda=0$ and find other solutions.

In summary, an NS-NS field strength supported on $S^1 _{\beta}\times S^2$, a constant radion,  a constant dilaton and a flat spatial metric in the remaining $d-2$ dimensions are a solution to all EOM.

\subsubsection{A solution with the geometry of $S_{\beta} ^1 \times S^2\times \mathbb{R}_{\Phi_d}  \times \mathbb{R}^{d-3}$ and a linear dilaton}

Here we consider the case of a linear dilaton in some specific direction, which we denote by $x$,
\begin{equation}
\Phi_d (x) = \Phi_0 +Qx.
\end{equation}
Substituting the flux solution from Eq.~(\ref{Hsquared}) and the Ricci scalar of the two-sphere from Eq.~(\ref{Ricci}) with $r_0=\frac{\sqrt{\alpha'}}{2\sqrt{\epsilon}}$  from Eq.~(\ref{Sphere2b}) into the dilaton EOM (\ref{dilatonR}) implies that
\begin{equation}
Q =\pm \frac{\sqrt{\epsilon}}{\sqrt{\alpha'}}+O(\epsilon).
\end{equation}

A standard issue when discussing linear dilaton backgrounds is the appearance of a strong coupling region. The string coupling $g_s$ scales as $g_s\sim e^{\Phi_0+Qx}$, thus in some region of $x$ it becomes large. A potential fix to this issue is that higher-order terms in the string coupling do not allow it to become strong.

In summary, we established that that an NS-NS field strength on $S^1 _{\beta}\times S^2$, a constant radion, a linear dilaton in one of the spatial dimensions and a flat spatial metric in $d-3$ dimensions constitute a consistent solution to all the EOM.

\subsection{Solutions with a Ramond-Ramond flux and the geometry of $S^1 _{\beta} \times S^2 \times S^{p-1}\times AdS_{d-p-1}$}

In this subsection we discuss solutions with the geometry of
$S^1 _{\beta} \times S^2 \times S^{p-1}\times AdS_{d-p-1}$  and constant dilaton. Here the $AdS_{d-p-1}$ factor is Euclidean.
As for the cases discussed previously,
\begin{equation}
H_{\tau \mu \nu} = \tilde{h} \epsilon_{\tau \mu \nu},
\end{equation}
where $\tilde{h}$ may not necessarily be equal to $h$ from the previous subsection.
The RR flux is supported on $S^1 _{\beta}\times S^{p-1}$,
\begin{equation}
F_{\tau \mu_2 ... \mu_{p-1}} = f \epsilon_{\tau \mu_2 ... \mu_{p-1}}.
\end{equation}
We consider $p=3,5$ for type IIB and $p=4$ for type IIA. For $p=1$ we find that flux quantization is inconsistent with the $\sigma$ EOM. For $p=2$, the combined spatial metric and dilaton equation in the $S^1$ direction is not satisfied.
As will be shown, the combination of the two kinds of fluxes fixes the dilaton in addition to stabilizing the radion.

For a constant dilaton, Eq.~(\ref{dilatonR}) implies that
\begin{equation}
\label{RhR}
R = \frac{1}{2} \tilde{h}^2 + O(\epsilon^2).
\end{equation}
From equation~(\ref{SigmaR}) it follows that
\begin{equation}
\label{fheps}
\frac{1}{4} e^{2\Phi_d} f^2 + \frac{1}{2} \tilde{h}^2 = \frac{4\epsilon}{\alpha'}.
\end{equation}
We now turn to equations (\ref{EinsteinR}).
The spatial metric equation on $S^2$ requires that
\begin{equation}\label{S2}
\frac{1}{r_0 ^2}  -\frac{1}{2}\tilde{h}^2 +\frac{1}{4}e^{2\Phi_d} f^2 =0,
\end{equation}
and similarly for the spatial metric on $S^{p-1}$ with radius $r_1$,
\begin{equation}\label{Sphere}
\frac{p-2}{r_1 ^2} -\frac{1}{4} e^{2\Phi_d} f^2 =0.
\end{equation}
For the $AdS_{d-p-1}$ factor, we find that
\begin{equation}\label{AdS}
-\frac{d-p-2}{l_{AdS}^2} +\frac{1}{4} e^{2\Phi_d} f^2=0.
\end{equation}
From the trace of the d-dimensional metric-dilaton Eq. (\ref{EinsteinR}) we obtain
\begin{equation}
R-\tilde{h}^2 -\frac{p-1}{2} e^{2\Phi_d} f^2 + \frac{d}{4} e^{2\Phi_d} f^2=0.
\end{equation}
Using Eq.~(\ref{RhR}), it follows that
\begin{equation}
\tilde{h}^2 =\frac{1}{2} e^{2\Phi_d} f^2 \left(d-2p+2\right),
\end{equation}
and from Eq. (\ref{fheps}) we obtain our final results for $f$ and $\tilde{h}$,
\begin{equation}
f= e^{-\Phi_d} \frac{4}{\sqrt{d-2p+3}}\sqrt{\frac{\epsilon}{\alpha'}},
\end{equation}
\begin{equation}
\tilde{h} = \sqrt{\frac{2(d-2p+2)}{(d-2p+3)}}\sqrt{\frac{\epsilon}{\alpha'}}.
\end{equation}
The scales of the spheres and the AdS space are determined from Eqs.~(\ref{S2}), (\ref{Sphere}) and (\ref{AdS}):
The radius of $S^2$ is given by
\begin{equation}
r_0= \sqrt{\frac{d-2p+3}{d-2p-2}}\sqrt{\frac{\alpha'}{\epsilon}},
\end{equation}
the radius of $S^{p-1}$ is given by
\begin{equation}
 r_1  = \sqrt{\frac{(p-2)(d-2p+3)}{4}}\sqrt{\frac{\alpha'}{\epsilon}},
 \end{equation}
and the AdS scale by
\begin{equation}
 l_{AdS} = \sqrt{\frac{(d-p-2)(d-2p+3)}{4}}\sqrt{\frac{\alpha'}{\epsilon}}.
\end{equation}
The $H_3$ flux is quantized, implying that $\epsilon \propto \frac{1}{n_H^2}$ with $n_H\in \mathbb{Z}$. The quantization of $F_p$ constraints the string coupling $e^{\Phi_d}\propto \frac{ n_H^{p-2}}{n_F}$ where $n_F\in \mathbb{Z}$. Since the solution relies on weakly-coupled EFT with light winding modes, the relevant region in parameter space is $1\ll n_H $ and $n_H^{p-2}\ll n_F$.

In summary, an NS-NS field strength on $S^1 _{\beta}\times S^2$, RR flux on $S^1 _{\beta}\times S^{p-1}$, a constant radion,  a constant dilaton and an AdS space in the remaining ${d-p-1}$ dimensions, is a consistent solution to all the EOM.

\subsection{Aspects of stability}

Here we only discuss some aspects of the perturbative stability of the solutions, deferring a full stability analysis to a future publication.

First, we point out that the form of the effective potential for the winding modes guarantees stability relative to fluctuations about $\chi \chi^*= \frac{\epsilon}{\kappa^2}$.
Second, the integrated fluxes are perturbatively protected from instabilities. One can, for instance, perturb both the $H_3$ flux, the $F_p$ flux and the radion $\sigma$ in a correlated manner, but we argue that the solutions are stable against this. Indeed, consider the terms in the action in which $\sigma$ appears, where in particular the $|H_3|^2,|F_p|^2$ terms are positive,
\begin{eqnarray}
&S[\sigma] = \beta \int  d^d x \sqrt{G} e^{-2\Phi_d}\left[ \frac{2}{\alpha'}\chi \chi^*e^{2\sigma} +\frac{1}{8\kappa^2}e^{-2\sigma} H_{\tau \mu \nu}H_{\tau} ^{~ \mu \nu} \right. \cr  &+ \left. \frac{1}{4\kappa^2 (p-1)!} e^{2\Phi_d-\sigma} F_{\tau \mu_2...\mu_p} F_{\tau} ^{~\mu_2...\mu_p}\right].
\end{eqnarray}
For all the solutions discussed above, the extremum for the field $\sigma$ is a minimum. Expanding about the solution, the mass-squared of $\sigma$ scales as $\frac{\epsilon}{\alpha'}$.

 Further investigation is required to complete a full stability analysis.

\subsection{Entropy}

The entropy density carried by the winding modes is related to the Lagrangian density of the winding modes $\mathcal{L}_{\chi}$ in the $d$-dimensions through
\begin{equation}
s = \left(\beta \partial_{\beta} -1\right) \mathcal{L}_{\chi} = \frac{2\beta^2 e^{-2\Phi_d+2\sigma}}{(2\pi \alpha')^2} \chi \chi^*.
\end{equation}
Setting $\beta e^{\sigma} \approx \beta_H = 2\pi \sqrt{2\alpha'}$, $\chi \chi^* = \frac{\epsilon}{\kappa^2}$ and $\kappa^2 = 8\pi G_N$, one obtains
\begin{equation}
s =\frac{\epsilon}{2\pi \alpha' G_N}.
\end{equation}
The scaling of the entropy as $G_N ^{-1}$ characterizes the classical entropy of winding-mode condensates  \cite{MaldacenaLargeD},\cite{CMW},\cite{EntropyCondensate},\cite{Amit},  \cite{WindingBackreaction} (for an earlier work, see \cite{Dabholkar}).

\section{Worldsheet Field Theories}

In this section we write a classical action for some worldsheet (WS) field theories, and propose that they have fixed points corresponding to the backgrounds in Sect. 3.2. The WS theories describe a string propagating in the thermal target-space background. The analysis below is done to leading order in $\alpha'$. It would be interesting to validate our proposal by studying the renormalization group flow of the couplings in the WS action. 
Also it would be interesting to extend the leading-order result to higher-orders in $\alpha'$, and perhaps even find an exact CFT.

The spectrum of the WS of type II consists of $X^{\mu}$ bosonic fields and the WS fermions $\psi^{\mu},\tilde{\psi} ^{\mu}$.
The WS CFT action is given by
\begin{equation}
S_{WS} = S_X + S_{\psi,\tilde{\psi}} + S_{ghosts} + S_{SG}.
\end{equation}
\begin{equation}
S_X =\frac{1}{2\pi \alpha'}\int d^2 z \left(e^{-2\sigma} \partial \tau\bar{\partial} \tau+ G^{\mu \nu}\partial X_{\mu} \bar{\partial} X_{\nu} +B^{\mu \nu} (X) \partial X_{\mu} \bar{\partial} X_{\nu}+\alpha' \mathcal{R} \Phi_D(X) \right).
\end{equation}
\begin{eqnarray}
&S_{\psi,\tilde{\psi}}= \frac{1}{4\pi} \int d^2 z \left(e^{-2\sigma} \psi_{\tau} \bar{\partial} \psi_{\tau}+ G^{\mu \nu} \psi_{\mu} \bar{\partial} \psi_{\nu}\right)+\nonumber \\
&\frac{1}{4\pi} \int d^2 z \left(e^{-2\sigma} \tilde{\psi}_{\tau} \partial \tilde{\psi}_{\tau}+ G^{\mu \nu} \tilde{\psi}_{\mu} \partial \tilde{\psi}_{\nu}\right).
\end{eqnarray}
So far the action is standard; the target space fields in this action are given by:
\begin{eqnarray}
G_{\mu \nu} (X) &=& \delta_{\mu \nu}, \ \ \ \mu,\nu \neq 1,2, \\
G_{ij}&=&\text{diag} (1,\sin^2 (\theta)), \ \ i,j=1,2, \\
\sigma(X) &=& -\epsilon, \\
B_{\alpha \beta}(X)&=&0, ~\alpha,\beta\neq \tau, \\
B_{\tau i} (X) &=& \frac{1}{2} h \epsilon _{\tau ij} X^{j}, ~i,j=1,2.
\end{eqnarray}
The dilaton is either a constant or linear:
\begin{eqnarray}
\Phi_d(X) &=& \Phi_0, \\
\Phi_d (X) &=& \Phi_0 + Q x~,~ Q^2 = \frac{\epsilon}{\alpha'}.
\end{eqnarray}
We deform the action by adding the following term, a sum of the winding vertex operators\footnote{We focus on the real positive solution $\chi = \frac{\sqrt{\epsilon}}{\kappa}$.}:
\begin{equation}
\label{SG}
S_{\chi}=2\mu \int d^2 z~  \psi_{\tau} ~ \tilde{\psi}_{\tau}~ \cos\left[\frac{\beta e^{\sigma}}{2\pi \alpha'}\left(\tau_L-\tau_R\right) \right].
\end{equation}
Here,
\begin{equation}
\mu\equiv \sqrt{\epsilon}.
\end{equation}
In \cite{HagedornEFT} we showed that the ``Sine-Gordon'' term in Eq. (\ref{SG}) reproduces the EFT amplitudes. One can compute the leading-order conformal weight of the deformation, $h=1+O(\epsilon)$. In the previous section we verified that the equations of motion are satisfied to leading order in the $\alpha'$ expansion. This leads us to expect that the field theories with the above action have fixed points corresponding to the backgrounds described in Sect. 3.2. The $\alpha'$ corrections are small because space gradients of any of the target-space fields either vanishes (for instance $\chi,\chi^*, G_{\mu \nu}$) or scale as $\sqrt{\frac{\epsilon}{\alpha'}}$.

The Sine-Liouville CFT and the $\mathcal{N}=2$  supersymmetric Liouville theory include a term similar to $S_{SG}$ in Eq.~(\ref{SG}). However, in our case the winding-mode vertex operators are uniform in space, while in the Sine-Liouville theory (and in its supersymmetric version) they are multiplied by $e^{-\frac{\phi}{Q}}$, where $\phi$ is the dimension in which the dilaton is linear. Also, in our case,  $\mu$ is a small parameter in contrast to the Sine-Liouville theory, which has ${1}/{\mu}$ to a positive power, as its small expansion parameter.

\section{Comparison with the Cigar Background}

The purpose of this section is to compare and contrast the thermal equilibrium background that we found and the $SL(2,R)_k/U(1)$ cigar and cylinder backgrounds.

The asymptotic radius of these backgrounds is related to the level $k$ by $\sqrt{\alpha'k}$, they are weakly-curved for large $k$ and for the cigar, the cycle vanishes in size at the tip. The geometry and dilaton in the cigar are given by:
\begin{equation}
ds^2 =\tanh^2 \left(\frac{\rho}{\sqrt{k \alpha'}}\right) d\tau^2+ d\rho^2, ~\tau \sim \tau +2\pi \sqrt{\alpha' k}~,
\end{equation}
\begin{equation}
\Phi(\rho) = \Phi_0- \log\left(\cosh\left(\frac{\rho}{\sqrt{k\alpha'}}\right)\right).
\end{equation}
In addition, a negative CC is required,
\begin{equation}
\Lambda= -\frac{2}{\alpha' k}.
\end{equation}
The gauged Wess-Zumino-Witten CFT describes the target space to all orders in the $\alpha'$ expansion \cite{EdCigar}. The Fateev-Zamolodchikov-Zamolodchikov duality \cite{Zamolodchikov} can be used to describe the cigar theory in terms of the Sine-Liouville CFT on a cylinder \cite{KKK}. A supersymmetric version of this CFT is the $\mathcal{N}=2$ Liouville CFT  \cite{LST},\cite{N2Liouville}.

The cigar background includes a winding-mode condensate \cite{Kutasov1}, \cite{Sunny1},\cite{Sunny2}. 
Neglecting $\alpha'$ corrections and treating the winding-mode as a fluctuation about the background, its profile is given by
\begin{equation}
\chi (\rho )\propto \frac{1}{\cosh^k \left(\frac{\rho}{\sqrt{\alpha' k}}\right)}.
\end{equation}
This is a zero-mode solution of the EOM of the winding modes. 
The backreaction of the winding-mode condensate on the geometry was discussed recently in \cite{WindingBackreaction}, where it was shown that the profile of the winding-mode condensate is determined by the Nambu-Goto action of a string wrapping the cigar up to a radial distance $\rho$:
\begin{equation}
\chi(\rho) =e^{-\frac{\gamma}{2}} e^{-\frac{\beta}{2\pi \alpha'} \int^{\rho} e^{\sigma(y)}dy},
\end{equation}
where $\gamma$ is the Euler-Mascheroni constant.
In contrast to the space-filling condensate for the thermal equilibrium solution, for the cigar and for large $k$, the winding modes are localized near the tip of the cigar. One can thus focus on the region near the tip to make the comparison. However, the important difference between the thermal equilibrium backgrounds and the cigar background is in the nature of the thermal cycle: whether it is contractible or not. Additionally, the cigar is not uniform while our solutions with constant dilaton are uniform.

Another target-space solution is that of the $\mathcal{N}=2$ Liouville theory with $k$ slightly smaller than 1, corresponding to an effective temperature slightly higher than the Hagedorn temperature. This solution differs from the thermal equilibrium solution in that it does not include fluxes.

\section{Comparison with the Horowitz-Polchinski Background}

The purpose of this section is to compare and contrast the thermal equilibrium background and the HP background \cite{HP}, whose properties were reviewed recently in \cite{CMW}.
The HP background exists for $d\leq 5$ and has the topology $S_1 ^{\beta} \times \mathbb{R}^{d}$.
The asymptotic radius of the thermal circle corresponds to a temperature close to, but below, the Hagedorn temperature. The winding-mode condensate is localized in a of a region in space of radius $\sqrt{\frac{\alpha'}{|\epsilon|}}$ and its magnitude is of the order of $|\epsilon|$. In our convention, used in Eq.~(\ref{StartingPoint}), the magnitude also scales with the inverse of the string coupling $\frac{1}{g_s}$. The thermal circle becomes slightly smaller than the asymptotic circumference, and the local temperature surpasses the Hagedorn temperature in the central region where the condensate is localized. A Lorentzian interpretation of the Euclidean condensate is that of a gas of highly excited strings under the influence of Newtonian gravity.

The entropy density is proportional to $|\chi|^2$ which is of order of $\frac{|\epsilon|^2}{G_N}$. This scaling should be compared to the thermal equilibrium solutions for which the thermal circle is constant and $|\chi|^2\sim\frac{\epsilon}{G_N}$.
Another difference between the two solutions is the importance of the winding self-interaction quartic coupling term relative to the coupling of the radion and two winding modes. For the HP solution, the quartic coupling is suppressed by a factor of $|\epsilon|$, whereas in the thermal equilibrium solution, both are of order $\epsilon^2$.
In addition, the HP solution is not uniform due to the condensate, while our solutions are uniform.

In Table 1, we summarize  the properties of the of the thermal equilibrium backgrounds compared to those of the $SL(2,R)/U(1)$ backgrounds and to the HP background.
\begin{table}[h]
\begin{center}
\resizebox{15cm}{!}{
\begin{tabular}{ |l |l |l|l|l| }
	\hline
	 &$ S^1 _{\beta}\times S^2 \times {\cal M}^{d-2} $ &  $SL(2,R)_{k\gg 1} / U(1)$& $SL(2,R)_{k\lessapprox 1} / U(1)$ & HP \\
	\hline
	Uniform & Yes, with const. Dilaton & No & No & No\\
	\hline
	Curvature& Weak & Weak (without backreaction)& Strong & Weak\\
	\hline
	Thermal equilibrium  & Yes & No & Yes& Approximate\\
	\hline
	 Asymptotic Temperature  &$\gtrapprox T_H$ &   $\ll T_H$ & $\gtrapprox T_H$ & $\lessapprox T_H$ \\
	\hline
	Dilaton, $\Phi_D$ &const./linear& at $~\rho\approx \sqrt{2\alpha'}$, $\sim$ const.& Linear   & $\approx~$ const.  \\
	\hline
	Winding condensate & Space filling &  Near the tip & Space filling &Size $\sim {1}/{\sqrt{|\epsilon|}}$\\
	\hline
	Flux  &  Yes & No & No & No\\
	\hline
\end{tabular} }
\caption{In this table, we list the similarities and differences between the various target spaces, the $S^1 _{\beta}\times S^2 \times {\cal M}^{d-2}$ background, the cigar background, the super-Liouville one and the HP background. We would like to emphasize that while BHs can be in thermal equilibrium with their environment, here we simply point out that the local temperature in Euclidean signature varies.}
\end{center}

\end{table}

\newpage
\section{Summary and Discussion}

In this paper, we derived the equations of motion for the light fields of closed strings above and near the Hagedorn temperature. We verified explicitly that the equations of motion possess several solutions with an isotropic and homogeneous flux keeping the thermal cycle fixed and stable. The solutions differ by the behavior of the dilaton - constant or linear, by whether the cosmological constant vanishes and whether they include a non-vanishing RR flux. The cosmological constant could perhaps correspond to the value of the minimum of the potential of some other target-space fields. It would be interesting to find out whether this is indeed the case.

We thus demonstrated that a state in thermal equilibrium in the Hagedorn phase of string theory, slightly above the Hagedorn temperature, does indeed exist. Typically, the condensate of light winding modes is interpreted as a collection of strings at high temperature. From this Lorentzian perspective, our Euclidean solutions provide descriptions of thermal bath of strings in the Hagedorn phase.

However, attempting to analytically continue the Euclidean solution to Lorentzian signature via the standard procedure $\tau \sim \tau +\beta e^{\sigma} \to it $, leads to several issues. For example, the continued three-form flux would fail to satisfy quantization, would be sourced by imaginary string sources and would have a negative energy. We interpret these issues as suggesting that a more appropriate continuation should involve the thermodynamical quantities, such as the entropy, energy and pressure, which are intrinsic properties of the thermal state.

As the temperature decreases, from above, to the Hagedorn temperature $T_H$, $\epsilon=\frac{T-T_H}{T_H}$ decreases to 0. One could imagine that as $\epsilon$ is decreased, the volume of the spheres increase, the values of the asymptotic fluxes (and the slope of the linear dilaton) decrease and formally vanish at the Hagedorn temperature. Decreasing $\epsilon$ further, we envision a transition to a thermal phase of short strings in which the winding-mode condensate vanishes.

We discussed aspects of the stability of the solutions, as well as the entropy carried by the winding modes.  The similarities and differences between the thermal background, the cigar background and the HP background were also discussed and elucidated.

We do not expect our solutions to be related to black holes. Our solutions are translation-invariant while the black-holes solutions are not. In \cite{CFT4dS},\cite{dSHigh}, evidence was presented that a simple version of the solution with constant dilaton and no RR flux is related to de-Sitter space. For instance, both are isotropic and uniform. The thermal state that we have found can serve a starting point for describing a string theory dual to asymptotically de Sitter space.

However, we think that our theory does have another solution (or multiple solutions) with  $T>T_H$ (here $T$ is the asymptotic temperature) in which the quartic interaction is important  and where the thermal cycle pinches off.  This solution should include a winding condensate and corresponds to black-hole solution(s). It would be interesting to investigate this further. For references discussing  relations between black holes and the super-Hagedorn phase, see for example \cite{Barbon:2004dd},\cite{Kruczenski:2005pj} and \cite{Horowitz:2006mr}.

\section*{Acknowledgements}
{We would like to thank Sunny Itzhaki for motivating us to start this investigation. We are grateful to Ofer Aharony, Yiming Chen, Amit Giveon, Sunny Itzhaki and David Kutasov for comments on the manuscript.  The work of R. B. and Y. Z. is supported by the German Research Foundation through a German-Israeli Project Cooperation (DIP) grant ``Holography and the Swampland.''
The research of Y. Z. is supported by the Adams fellowship.}

\section*{Appendix - Winding Current and the B-Field}
An S-matrix calculation of the amplitude of a massless modes with polarization tensor $\epsilon_{3\mu \nu}$, with two winding modes of winding numbers $\pm 1$ yields \cite{HagedornEFT}
\begin{equation}
	\mathcal{S} (k_1,k_2,k_3) = -\frac{1}{2} \kappa' \beta (2\pi)^d \delta^d \left(k_1 ^{\perp}+ k_2 ^{\perp}+k_3 ^{\perp}\right) \epsilon_{3\mu \nu}k_{12 } ^{\mu} k_{12} ^{\nu} ~,~ k_{12}\equiv k_1 - k_2.
\end{equation}
Below it is explained how to extract a term in the effective action which reproduces this amplitude.
As always, the $(2\pi) \delta^d (...)$ factor transforms into $\int d^d x$ in the effective action. For $\chi,\chi^*$ and $B_{\tau \mu}$ one should take $k_1 ^{\tau} = \frac{\beta}{2\pi \alpha'},k_2 ^{\tau} = -\frac{\beta}{2\pi \alpha'}$. In addition, factors of $-i\partial_{\mu}$ are induced by the spatial momenta of the winding modes. One should also take into account a relative minus sign between the effective field theory amplitude and the string amplitude in Euclidean signature. These steps yield the following term in the d-dimensional effective action
\begin{equation}
	i\frac{\beta}{2\pi \alpha'}\kappa' \int d^d x  B_{\tau} ^{~\mu}\left(\chi \partial_{\mu} \chi ^* - \chi^* \partial_{\mu} \chi\right) . 
\end{equation}
Applying the field redefintions which appear between Eqs. (57)-(58) of \cite{HagedornEFT}, $\chi \to \sqrt{\beta}\chi ~,~ \chi^* \to \sqrt{\beta} \chi^*$ and $B\to \kappa' B$, one eventually obtains
\begin{equation}
	i\frac{\beta}{2\pi \alpha'}\beta \int d^d x~ e^{-2\Phi} \sqrt{G} B_{\tau} ^{~\mu}\left(\chi \partial_{\mu} \chi ^* - \chi^* \partial_{\mu} \chi\right) . %
\end{equation}
We have also included the dilaton and metric factors to agree with the standard form of the covariant effective action at leading order in the string coupling. This result coincides  with Eq. (4.6) of \cite{Troost} who analyzed the Heterotic string in Lorentzian signature.

\end{document}